\documentclass[prb,aps,twocolumn,superscriptaddress,showpacs]{revtex4}

\usepackage{graphicx}

\begin{document}

\title{Direct observation of the near-surface layer in Pb(Mg$_{1/3}$Nb$_{2/3}$)O$_{3}$ using neutron diffraction}

\author{K. Conlon}
\affiliation{Department of Materials Science, McMaster University, Hamilton, Ontario, L8S 4L7}

\author{H. Luo}
\affiliation{Shanghai Institute of Ceramics, Chinese Academy of Sciences, Shanghai, China, 201800}

\author{D. Viehland}
\affiliation{Department of Materials Science and Engineering, Virginia Tech., Blacksburg, Virginia, 24061}

\author{J. F. Li}
\affiliation{Department of Materials Science and Engineering, Virginia Tech., Blacksburg, Virginia, 24061}

\author{T. Whan}
\affiliation{National Research Council, Chalk River, Ontario, Canada, K0J 
1J0}

\author{J.H. Fox}
\affiliation{National Research Council, Chalk River, Ontario, Canada, K0J 
1J0}

\author{C. Stock}
\affiliation{Department of Physics, University of Toronto, Ontario, Canada 
M5S 1A7}

\author{G. Shirane}
\affiliation{Physics Department, Brookhaven National Laboratory, Upton, 
New York 11973}

\date{\today}

\begin{abstract}

Spatially resolved neutron diffraction as a function of crystal depth in Pb(Mg$_{1/3}$Nb$_{2/3}$)O$_{3}$ reveals the presence of a distinct near-surface region.  A dramatic change in both the lattice constant and the Bragg peak intensity as a function of crystal depth is observed to occur in this region over a length scale $\sim$ 100 $\mu$m.  This confirms a previous assertion, based on a comparison between high-energy x-rays and neutrons, that such a near surface region exists in the relaxors.  Consequences to both single crystal and powder diffraction measurements and previous bulk neutron diffraction measurements on large single crystals are discussed.

\end{abstract}


\maketitle

	The relaxor ferroelectrics, with the general formula Pb$B$O$_{3}$, have generated much interest recently due to their promising applications as piezoelectric devices.~\cite{Ye98:81,Park97:82}  Pb(Mg$_{1/3}$Nb$_{2/3}$)O$_{3}$ (PMN) and  Pb(Zn$_{1/3}$Nb$_{2/3}$)O$_{3}$ (PZN) are prototypical relaxor systems which display a diffuse transition with a broad frequency dependent dielectric response.  Despite much work on these materials there is no consensus on the low-temperature structure or ground state. 

\begin{figure}[t]
\includegraphics[width=10cm,bbllx=10,bblly=300,bburx=600,
  bbury=750,angle=0,clip=] {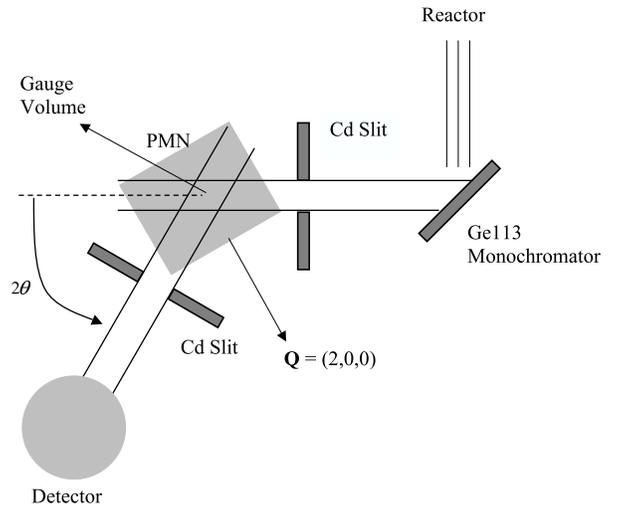}
\caption{\label{figure1} Schematic of the experiment used to measure spatially resolved strain with neutrons.  The mosaic of the Germanium monochromator was $\sim$ 15 $'$, and a single channel from the monochromator to incident slit defined a collimation of $\sim$ 30$'$.  A single slit was used on the scattered side.  The slits on both the incident and scattered sides were 300 $\mu$m wide and 5 mm high.}
\end{figure}

	Recent x-ray diffraction measurements by Xu \textit{et al.}, using both 32 keV and 67 keV x-rays, found the low temperature unit cell near the surface (probed by 32 keV) to be rhombohedral where the bulk (studied with 67 keV) was found to be cubic in shape.~\cite{Xu03:67,Lebon02:14}  Later, a neutron elastic scattering study on PZN found the Bragg peaks broadened considerably below T$_{c}$ both along the transverse and longitudinal directions, but no sign of any clear rhombohedral distortion was observed.~\cite{Stock04:69}  The neutron and x-ray results were reconciled by the assumption of a near-surface region with a rhombohedral unit cell.  A similar explanation was used to reconcile apparently contradictory results in PMN doped with 10\% PbTiO$_{3}$.~\cite{Gehring03:4289}  In this material, low-energy x-rays have been used to measure a rhombohedral unit cell while neutron diffraction results have pointed to a bulk cubic unit cell.  However, radial scans through the (1,1,1) Bragg peak, using neutron diffraction, found evidence for, at least, a segment of the sample having a rhombohedral unit cell.  A rhombohedral skin was, again, postulated based on a comparison of neutron and x-ray data.  Since neutrons are highly penetrating, such a near surface region must be much larger than several unit cells for there to be any effect in a neutron diffraction experiment.  Such a near-surface layer would also be expected to induce a large strain on the crystal.   

	The importance of a near surface layer with a different unit cell shape than the bulk is highlighted by powder diffraction work.  X-ray and neutron powder diffraction studies of the low-temperature structure of both PMN and PZN have used a two-phase model to describe the intensity profiles.~\cite{Iwase99:60,Mathan91:3}   One phase was found to be, on average, cubic while the second phase was rhombohedral.  An important question concerns whether or not the rhombohedral and cubic phases can be separated into a bulk and near-surface region as postulated to describe the neutron and x-ray single crystal data.  The near-surface region may play an important role in the thermal expansion properties of relaxors.  This has been noted in a recent x-ray experiment on unpoled PZN which measured the bulk (probed by 67 keV x-rays) to have comparably little thermal expansion with respect to the surface region (studied with 10 keV).~\cite{Xu04:3369}  Given the many unusual structural properties of the relaxors, it is clearly important to determine if there exists a macroscopic variation in structural properties.   

\begin{figure}[t]
\includegraphics[width=8cm] {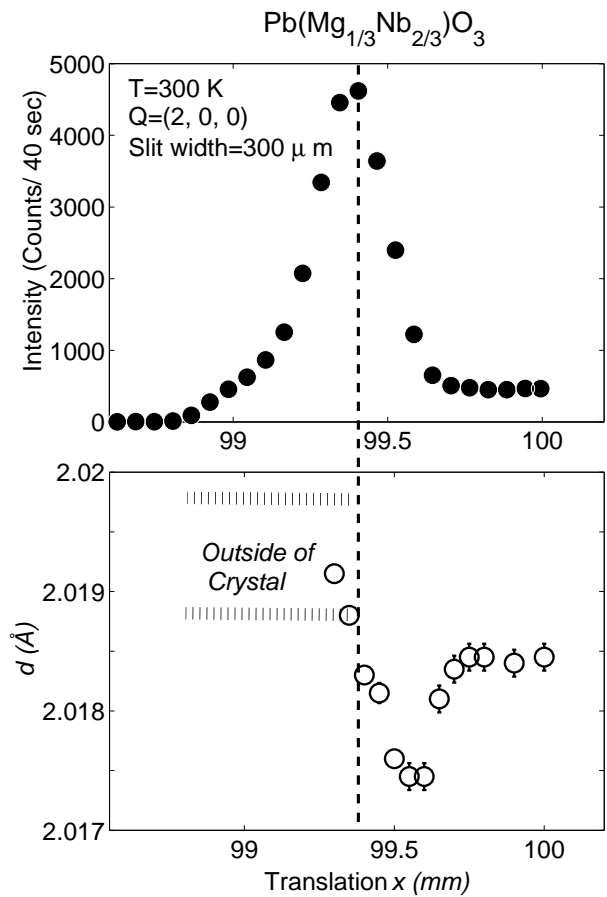}
\caption{\label{figure3} The upper panel plots the (2, 0, 0) Bragg peak intensity as a function of translation.  The lower panel displays the lattice constant (and hence the strain) as a function of distance into the sample.  A large $\sim$ 100 $\mu$m surface layer is observed based on a large variation of the lattice constant.  The vertical dashed line indicates the position of the sample surface.}
\end{figure}

\begin{figure}[t]
\includegraphics[width=8cm] {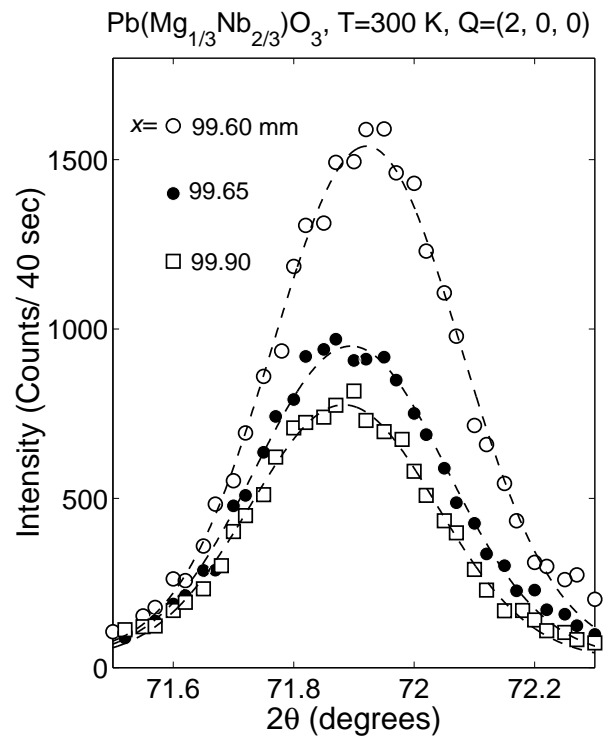}
\caption{\label{figure2} Radial $\theta-2\theta$ scans through the (2, 0, 0) Bragg peak for various translation values.  The decrease in intensity is due to extinction and a noticeable shift in the $2\theta$ position is clearly seen.  The dashed lines are fits to a Gaussian lineshape.}
\end{figure}

	To reconcile previous neutron and x-ray measurements it is very important to establish the existence of a surface layer with one technique (or instrument) to avoid systematic errors.  To study the possibility of such a near-surface region in relaxors we have investigated the strain as a function of crystal depth using a diffraction technique commonly applied to study spatially resolved strain in industrial materials.  The lattice constant and Bragg intensity is observed to vary strongly with penetration depth, which is suggestive of a significant near-surface region.  A large extinction effect is also observed which suggests the near-surface regions makes a significant contribution to the total scattering in single crystal neutron experiments.

	Neutron diffraction measurements were conducted on the L3 double-axis spectrometer located in the NRU reactor at Chalk River Laboratories.  A Germanium (113) crystal was used to produce a monochromatic incident beam with energy set to E$_{i}$=14.5 meV.  Higher-order neutrons were removed by placing a Pyrolytic Graphite filter in the incident beam.  The horizontal collimations were set to 60$'$-30$'$-\textit{S}-30$'$. The sample consisted of a 9.3 cc crystal of PMN, grown by the modified Bridgeman technique previously described.~\cite{Luo00:39, Luo03:xx}  The lattice parameter in the bulk was measured to be $\sim$ 4.04 \AA.  A large [100] cut surface was used to define the surface of the sample from which the (200) Bragg peak was studied in reflection geometry.  The PMN sample used for these experiments is the same crystal used in a recent phonon study.~\cite{Stock04:xx}

\begin{figure}[t]
\includegraphics[width=8cm] {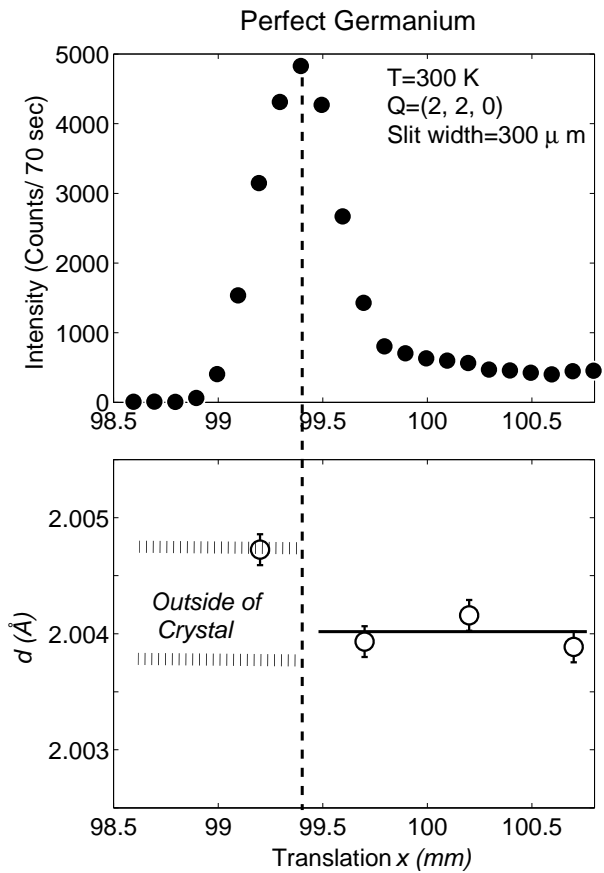}
\caption{\label{figure5} The upper panel plots the Germanium (2, 2, 0) Bragg peak intensity as a function of translation.  The lower panel displays the lattice constant (and hence the strain) as a function of distance into the sample. Unlike PMN, no significant strain is measured in the near-surface region.  The vertical dashed line indicates the position of the sample surface.}
\end{figure}

	Figure \ref{figure1} schematically illustrates the technique used to study the strain as a function of depth.  The incident and diffracted beams, defined by masks made of absorbing Cadmium, intersect in a small region of the sample denoted as the gauge volume.  The scattering geometry is analogous to the reflection technique commonly used in low-energy x-ray measurements.  In this experiment the incident and scattered slits were 0.3 mm wide and 5 mm tall.  The sample was positioned and aligned to a precision of $\sim$ 0.05 mm.  For our experiments, we have defined the translation axis to be along the [100] direction.  All measurements described here were conducted at room temperature.  This technique is very similar to that previously used to investigate the second length-scale phenomena in Tb and SrTiO$_{3}$.~\cite{Gehring95:51,Gehring93:71, Hirota94:49, Cowley78:11}

	There are several well characterized systematic effects that complicate near-surface strain scanning.  These have been described in detail elsewhere (Refs. \onlinecite{Webster96:3} and \onlinecite{Wang98:30}) and only result in strong systematic errors when the gauge volume is only partially submerged in the sample.  In the results presented here, we have ignored all data where the gauge volume is not completely immersed in the sample.  To further test for any near-surface experimental artifacts we have measured the strain near the surface in a perfect Germanium crystal using the (220) reflection.  The d-spacings for Germanium (220) is nearly identical to that of PMN (200) and therefore allows a direct test by keeping all the spectrometer angles the same (and hence the resolution function) while only changing the sample.  As noted later, this test resulted in no near-surface anomalies therefore ruling out instrument effects.

	To determine the exact location of the surface, translation scans were conducted at the (200) Bragg position.  The upper panel of Fig. \ref{figure3} plots the intensity as a function of translation distance.  The intensity as a function of depth cannot be described by a linear attenuation alone and therefore is strongly suggestive that primary extinction is the reason for the large (factor of 10) attenuation in the scattering intensity.  We note that a similar attenuation was observed in single crystals of Tb (as well as Germanium measured in this work) in a similar experiment and therefore such an effect is not unusual for single crystals.~\cite{Gehring95:51}  

	The precise surface position was taken to be the point of maximum intensity in this translation scan.  It is at this point that the gauge volume is entirely immersed in the sample while the effects of extinction are minimized.  On translation of the gauge volume into the sample, the intensity will be attenuated by the effects of extinction.  Rocking scans near the surface and in the bulk of the sample show identical rocking widths of 0.185 $\pm$ 0.007$^{\circ}$.  This indicates that any change in intensity or lattice constant is not likely the result of a variation of mosaic or the presence of another crystal domain near the surface.

	The lattice constant (and hence the strain) as a function of penetration depth is displayed in Fig. \ref{figure3}.  The lattice constant was extracted based on $\theta-2\theta$ scans through the (2, 0, 0) Bragg peak position.  Samples of these scans for three different translation positions are displayed in Fig. \ref{figure2}.  A large strain over a distance of $\sim$ 300 $\mu$m is observed.  Data for translation values outside of the crystal are unreliable due to the fact that the beam is not entirely immersed in the sample.  The fact that the width of the peak in the strain is close to the slit width used in the experiment indicates that peak width is limited by our resolution defined by the slit width.  Nevertheless, our result does suggest that the near-surface region is significant and of the order 100 $\mu$m in size and is not an effect of surface roughening over several unit cells.  

	Fig. \ref{figure5} plots the same measurement conducted with a perfect Germanium crystal using the (220) reflection.  Since the d-spacing for Germanium (220) is nearly identical to PMN (200), this provides a direct test to ensure that observed effect is not due to instrumental effects.  The Germanium data does not show any evidence of the lattice constant anomaly observed in PMN when the gauge volume is immersed in the sample.  Therefore, instrumental effects can be ruled out for the cause of the surface layer observed in PMN.   


	The presence of a significant skin also suggests that the two phases extracted from powder diffraction measurements may be the result of skin and bulk phases.  Our results suggest that the measured structural properties depend on the geometry and size of the sample, analogous to the case in SrTiO$_{3}$.~\cite{Hunnefeld02:66}

	The significant strain in the near-surface region may also reconcile apparent discrepancies between x-ray and neutron diffraction measurements of single crystal PZN and PMN.  Despite the fact that high-energy x-ray measurements found resolution limited peaks at low-temperatures, neutron diffraction found the Bragg peaks to be broad both along the transverse and radial directions.  The presence of a significant strain in a large surface layer may be the origin for the radial broadening observed in neutrons which sample the entire specimen volume.

	In conclusion, we have used a spatially resolved strain scanning technique to show the existence of a significant near-surface region in the relaxor PMN.  Its presence validates previous x-ray diffraction results which postulated such a region based on penetration depth arguments.  Our result also suggests that the two-phase models found to describe powder data may be separated into a surface and bulk component.  Even though our results clearly show the presence of a strong surface component, its role in the relaxor transition is not clear.  Further measurements are currently being conducted to study the dependence of the strain on PbTiO$_{3}$ doping as well as temperature.

\begin{acknowledgements}

We are grateful to P. Gehring, I. Swainson, and G. Xu for stimulating discussions and to L.E. McEwan, M.M. Potter, and R.Sabatini for invaluable technical help.  The work at the University of Toronto was supported by the Natural Science and Engineering Research Council and the National Research Council of Canada.  We also acknowledge financial support from the U.S. DOE under contract No. DE-AC02-98CH10886, and the Office of Naval Research under Grant No. N00014-99-1-0738.   

\end{acknowledgements}



\thebibliography{}

\bibitem{Ye98:81} Z.-G. Ye, \textit{Key Engineering Materials Vols. 155-156}, 81 (1998).

\bibitem{Park97:82} S.-E. Park and T.R Shrout, J. Appl. Phys. {\bf{82}}, 1904 (1997).

\bibitem{Xu03:67} G. Xu, Z. Zhong, Y. Bing, Z.-G. Ye, C. Stock, and G. Shirane, Phys. Rev. B {\bf{67}}, 104102 (2003).

\bibitem{Lebon02:14} A. Lebon, H. Dammark, G. Calvarin, and I.O. Ahmedou, J. Phys.: Condens. Matter {\bf{14}}, 7035 (2002).

\bibitem{Stock04:69} C. Stock, R.J. Birgeneau, S. Wakimoto, J.S. Gardner, W. Chen, Z.-G. Ye, and G. Shirane, Phys. Rev. B {\bf{69}}, 094104 (2004).

\bibitem{Gehring03:4289} P.M. Gehring, W. Chen, Z.-G. Ye, and G. Shirane, unpublished (cond-mat/0304289).

\bibitem{Xu04:3369} G. Xu, Z. Zhong, Y. Bing, Z.-G. Ye, C. Stock, G. Shirane, unpublished (cond-mat/0403369).

\bibitem{Westphal92:68} V. Westphal, W. Kleemann, and M.D. Glinchuk, Phys. Rev. Lett. {\bf{68}}, 847 (1992).

\bibitem{Iwase99:60} T. Iwase, H. Tazawa, K. Fujishiro, Y. Uesu, and Y. Yamada, J. Phys. Chem. Solids {\bf{60}}, 1419 (1999).

\bibitem{Mathan91:3} N. de Mathan, E. Husson, G. Calvarin, J.R. Gavarri, A.W. Hewat, and A. Morell, J. Phys.: Condens Matter {\bf{3}}, 8159 (1991).

\bibitem{Luo00:39} H. Luo, G. Xu, H. Xu, P. Wang and Z. Yin, Jpn. J. Appl. Phys. {\bf{39}}, 5581 (2000).

\bibitem{Luo03:xx} H. Luo, H. Xu, B. Fang, and Z. Yin, unpublished.

\bibitem{Stock04:xx} C. Stock, H. Luo, D. Viehland, J.F. Li, I. Swainson, R.J. Birgeneau, and G. Shirane, unpublished (cond-mat/0404086).

\bibitem{Gehring95:51} P.M. Gehring, K. Hirota, C.F. Majkrzak, and G. Shirane, Phys. Rev. B {\bf{51}}, 3234 (1995).

\bibitem{Gehring93:71} P.M. Gehring, K. Hirota, C.F. Majkrzak, and G. Shirane, Phys. Rev. Lett. {\bf{71}}, 1087 (1993).

\bibitem{Hirota94:49} K. Hirota, G. Shirane, P.M. Gehring, and C.F. Majkrzak, Phys. Rev. B {\bf{49}}, 11967 (1994).

\bibitem{Cowley78:11} R.A. Cowley, and G. Shirane, J. Phys. C {\bf{11}}, L939 (1978).


\bibitem{Webster96:3} P.J. Webster, G. Mills, X.D. Wang, W.P. Kang, and T.M. Holden, J. Neutron Research, {\bf{3}}, 223 (1996).

\bibitem{Wang98:30} X.-L. Wang, S. Spooner, and C. R. Hubbard, J. Appl. Cryst. {\bf{30}}, 52 (1998).

\bibitem{Hunnefeld02:66} H. Hunnefeld, T. Niemoller, J.R. Schneider, U. Rutt, S. Rodewald, J. Fleig, and G. Shirane, Phys. Rev. B {\bf{66}}, 014113 (2002).
\end{document}